\documentclass[twocolumn,amsmath,amssymb,prd,reprint,longbibliography,floatfix,8pt]{revtex4-1}

\usepackage{cancel}
\usepackage{enumitem}
\usepackage[utf8x]{inputenc}
\usepackage{graphicx}
\usepackage{dcolumn}
\usepackage{bm}
\usepackage[switch]{lineno}
\usepackage{float} 
\usepackage{subfigure}
\usepackage{tikz}
\usepackage{multirow}
\usetikzlibrary{arrows.meta}
\usetikzlibrary{arrows,decorations.pathmorphing,backgrounds,positioning,fit,petri}

\begin{document}

\title{Efficient Generation of Stable Linear Machine-Learning Force Fields \\ with Uncertainty-Aware Active Learning}

\author{Valerio Briganti}
\author{Alessandro Lunghi}
\email{lunghia@tcd.ie}
\affiliation{School of Physics, AMBER and CRANN Institute, Trinity College, Dublin 2, Ireland}

\begin{abstract}
{\bf Machine-learning force fields enable an accurate and universal description of the potential energy surface of molecules and materials on the basis of a training set of ab initio data. However, large-scale applications of these methods rest on the possibility to train accurate machine learning models with a small number of ab initio data. In this respect, active-learning strategies, where the training set is self-generated by the model itself, combined with linear machine-learning models are particularly promising. In this work, we explore an active-learning strategy based on linear regression and able to predict the model's uncertainty on predictions for molecular configurations not sampled by the training set, thus providing a straightforward recipe for the extension of the latter. We apply this strategy to the spectral neighbor analysis potential and show that only tens of ab initio simulations of atomic forces are required to generate stable force fields for room-temperature molecular dynamics at or close to chemical accuracy. Moreover, the method does not necessitate any conformational pre-sampling, thus requiring minimal user intervention and parametrization.}
\end{abstract}

\maketitle

\section*{Introduction}

Machine learning (ML) models for the generation of force fields (FFs) are becoming a prominent aid for researchers in different fields, including drug discovery\cite{Drug_discovery}, prediction of metastable structures\cite{metastable_structures}, heterogenous catalysis \cite{Platinum_hydrogen_Kozinsky}, and more\cite{Review_Unke,Review_Tkatch,Review_keith_Tkatchenko,Review_Musil_Ceriotti}.
In all these fields, ML permits to speed up calculations or to manage larger datasets, largely overcoming the problem of the computational costs inherent to electronic structure simulations. In recent years, many ML models for the generation of FFs have been presented, e.g. sGDML\cite{CHMIELA201938}, BP-NNP\cite{NN_Behler_Parrinello,DeepMD,ANI_Isayev}, GPR based models\cite{GAP,GPR_Ceriotti,GPR_Rasmussen}, PhysNet\cite{Physnet}, SchNet\cite{Schnet}, FCHL19 descriptors combined with different regressors\cite{FCHL19}, moment tensor potentials\cite{moment_tensor_potential}, message passage neural networks\cite{Cormorant,Nequip_Kozinsky, Schutt2017_Message_passage_neur_netw}, and many more. All these methods have been shown to be able to reproduce the potential energy surface (PES) of complex chemical systems with chemical or near-to-chemical accuracy. However, such incredible results often come with the burden of requiring a lot of electronic structure simulations to generate the necessary training data to reach high accuracy, often in the range of $10^3-10^6$ calculations\cite{Phase_change_neural_network,Graphite_diamond,gold_nanoparticles}. Such a scenario poses serious challenges to the widespread use of MLFFs.\\
 
Decreasing the size of the training set is a non-trivial challenge that depends on many different factors. Among the most crucial ones there is the complexity of the ML architecture used to map the PES and the approach used to select a training set. Although simple ML models, such as linear ones, achieve less accuracy than complex ones, they often perform better for small training sets in virtue of being less prone to over-fitting issues. In this work we will focus on this class of MLFFs and investigate the possibility to further optimize their generation in terms of accuracy and training set size. \\

A conventional way to learn the PES of a compound is to first perform ab initio molecular dynamics to sample a relevant number of configurations and their energy/forces\cite{NN_Behler_Parrinello,Phase_change_neural_network}. This approach can potentially achieve a good performance on both training and test sets, as the most statically relevant structures are automatically included. However, such approach does not guarantee that redundancies are not also included, potentially leading to large computational overheads. Moreover, the accurate representation of a molecular PES also requires the sampling of statistically-rare conformations, which by definition are not captured by small-size molecular dynamics samplings. Crucially, when such rare conformations are encountered during a molecular dynamics run, the MLFF must be able to correctly predict their energies and forces in order to avoid leading to unphysical scenarios and a breakdown of the system stability. This serious issue thus often requires a second step where additional configurations are sampled from a MLFF-driven MD run to achieve the desired stability.\\

Active learning (AL) strategies have a big potential to overcome these issues and lead to the generation of optimal training sets. Active learning is the process of iteratively selecting data to add to the training set, according to a user-determined criterion. Ideally, such criterion must be chosen in order to i) iteratively add configurations to the training set only if they significantly differ from the ones already included in the training set, thus avoiding unnecessary overheads and ii) include all and only configurations required to training the model. Even if conceptually simple, achieving an optimal active learning strategy is far from straightforward.

One of the most used AL approaches is the query by committee\cite{Committee_AL,Car_active_learning,Query_by_committee_AL,Less_is_more,Behler_tutorial,Physnet,Infrared_machine_learning}. In this method, multiple models are trained to learn the same training set, but with different sets of initial parameters, e.g. biases and weights in a neural network. For the same ML architecture, different models will generally perform similar predictions of energy and forces for molecular configurations similar to those sampled in the training set, but will widely differ if the information contained in the training set is not sufficient to extrapolate to new configurations. Therefore, the disagreement on the prediction of energies and/or forces among the committee of MLFFs is used to signal AL to stop and extend the training with a new configuration. 

Another common approach to AL is based on Bayesian uncertainty prediction and Gaussian Process Regression ML models\cite{GPR_Rasmussen,FLARE,Rupp_GP,Uteva_GP,De_Vita_GP}. Bayesian models are generally based on the idea of combining our prior beliefs on the phenomenon under study and observations to achieve predictive power for unlabeled inputs. One of the strengths of this class of methods is the built-in possibility to estimate uncertainties on predictions, thus leading to a straightforward implementation of active learning. 

At the best of our knowledge, only three implementations of AL methods have been proposed for linear MLFFs. Podryabinkin et al.\cite{PODRYABINKIN2017171} provided a mathematically rigorous definition of interpolation and extrapolation with respect to a given training set and proposed an AL strategy specifically tailored for linear ML models. This method requires defining a maximum degree of extrapolation that the regression can attempt without triggering the AL algorithm to act and has been successfully used to find new stable alloys and crystal structures\cite{Shapeev_crystal_structure,Shapeev_method_new_alloys}. Some of the present authors instead tested a Gaussian metric over atomic environments' fingerprints to measure the similarity of newly encountered environments with respect to the structures spanned by the training set and trigger AL accordingly when the dissimilarity is above a certain threshold\cite{Lunghi2019-ug}. Very recently, a linear ML model based on Atomic Cluster Expansion (ACE)\cite{Dratzl_ACE} has been implemented together with an active learning process that combines elements of query by committee and Bayesian uncertainty prediction\cite{Csany_HAL,ACE_active_learning_Lysogorski}. Despite the successful use of Bayesian regression to perform AL in the latter work, no details on its robustness and implementation details were provided.\\

Similarly to the philosophy of Gaussian Process Regression, here we use the theory of linear regression to estimate the uncertainty of a model over predictions, provide a unified picture of all these approaches recently appeared in literature, and benchmark the capability of these principles to form the basis of an AL method for linear MLFFs. 

We assess the validity of this AL workflow by benchmarking the performance of the spectral neighbour analysis potential (SNAP)\cite{SNAP} over learning the revised MD17 data set\cite{Christensen_2020}. Moreover, we apply our method to four molecules of growing complexity, including coordination compounds and open-shell systems, and demonstrate that the proposed protocol generates MLFFs able to withstand stable molecular dynamics at room temperature starting from only one configuration in the training set and requiring a small amount of ab initio training data. This strategy can be readily applied to other linear MLFFs and used to tackle a wide range of chemical systems.

\section*{Methods}

\subsection*{Spectral neighbor analysis potential}

The MLFF used in this work is SNAP \cite{SNAP}. This method is based on the expansion of the total energy of the system in a sum of single atomic contributions, which are further expanded in a linear combination of bispectrum components
\begin{equation}
    E=\sum^{N_i}_{i} E_i=\sum^{N_i}_{i}\sum^{N_k}_{k}c_{k}(\alpha_i)B_{k}(i) \:,
\label{SNAP_equation}    
\end{equation}
where $B_{k}(i)$ is the $k$-th bispectrum component of atom $i$, and provides a geometrical description of its atomic environment within a cutoff radius $R_{cut}$. $N_k$ and $N_i$ are the number of bispectrum components in the expansion and the number of atoms in the system, respectively.
The coefficients $c_{k}(\alpha_i)$ depend on the atom species identified by the index $\alpha_i$, which can take an integer value between 1 and $N_{species}$, where $N_{species}$ is the number of atomic species in the system. A corresponding definition of forces in terms of bispectrum components can be easily obtained by taking the derivative of Eq. \ref{SNAP_equation} with respect to the atomic positions.  The terms $B_{k}(i)$ and their derivatives with respect to atomic coordinates are calculated using LAMMPS\cite{LAMMPS}. For a dataset of geometries and energies/forces, Eq. \ref{SNAP_equation} can be written as
\begin{equation}
\textbf{Y}=\textbf{X}\textbf{c} \:,
\label{SNAP_vectorial}
\end{equation}
where \textbf{Y} is a $N_{data} \times 1$ vector containing the target quantities to reproduce, either values of forces or energies. Defining $M=N_{kinds} \times N_{k}$ with $N_{kinds}$ being the number of atomic species in the system, \textbf{X} is a $N_{data} \times M$ matrix encoding Eq. \ref{SNAP_equation}, whilst the vector \textbf{c} assembles the coefficients $c_{k}(\alpha_i)$.

The training of SNAP requires the minimization of the loss function
\begin{equation}
    \mathcal{L}(\textbf{Y},\textbf{c})= \frac{\left\| \textbf{Y}- \textbf{X}\textbf{c} \right\|^{2}}{2}+\frac{\lambda}{2}\textbf{c$^{T}$}\textbf{c}\:,
    \label{loss_function}
\end{equation}
where $\lambda$ is a regularization parameter. The coefficients that minimize the loss function are thus given by \cite{bishop2006pattern}
\begin{equation}
    \textbf{c}=(\lambda\textbf{I}+\textbf{X}^T\textbf{X})^{-1}\textbf{X}^T\textbf{Y}
\label{classical_result}
\end{equation}

\subsection*{Uncertainty-driven active learning}

The AL workflow requires the following steps:
\begin{itemize}
    \item generate SNAP with a starting training set;
    \item run molecular dynamics and evaluate the uncertainty on the target quantity of the FF (energy and/or forces) at each step. If the uncertainty on the structure is higher than a certain threshold, an \textit{ab initio} calculation is performed on the new structure and the newly available information is included in the training set and the model retrained. If the uncertainty is low enough, MD keeps running;
    \item repeat the first two steps until the model can terminate a full MD of the desired duration without finding new structures.
\end{itemize}

The design of a method able to estimate the uncertainty and the definition of a stopping criterion are the key aspects of this method. In the following we detail the proposed protocol for such quantities. \\

The method for the estimation of the uncertainty is based on the classical theory of statistics of the linear least squares method. Let us first address two variables in order to easily visual the method's working principle. In this case $Y$ and $x$ are related by a linear mapping\cite{rotondi2011probability}
\begin{equation}
    Y= a +bx+Z \:,
    \label{2d-force_field}
\end{equation}
where $a$ and $b$ are coefficients to be determined and $Z$ captures a random component to $Y$ for which we have chosen a capital letter in order to stress its statistical nature. Once the coefficients are determined, we can obtain a value of the prediction $\hat{y}$ for every value of $x$. Crucial to our study, we can associate an error to the fit parameters and by propagation an uncertainty in predictions. This represents the core concept by which we predict uncertainty. In the 2-variable case, the estimation of the variance on the prediction $\hat{y}$ is given by \cite{rotondi2011probability}
\begin{equation}
    s^2(\hat{a}+\hat{b}x+z)=s^2_z\left[1+\frac{1}{n} \left( 1+\frac{(x-\left\langle x \right\rangle)^2}{\bar{s}_x^2} \right) \right] \:,
\label{error_prediction}
\end{equation}
where $s_z$ is the standard deviation of $Z$ which is here approximated as the difference between training data and corresponding predictions, $n$ is the cardinality of the training set, $ \left\langle x  \right\rangle$ is the mean of the distribution of $x$ and $\bar{s}_x^2$ is the variance of the values of $x$.
Assuming the variables $Z$ and $Y$ to have a Gaussian distribution, it is possible to show that the interval with confidence level $CL=1-\alpha$ is given by
\begin{equation}
    y(x)\in \hat{y}\pm t_{1-\alpha}s(\hat{y}(x))\:,
    \label{confidence_interval}
\end{equation}
where $t_{1-\alpha}$ is the quantile of the \textit{t-Student} distribution with $n-2$ degrees of freedom. \\

\begin{figure}
      \includegraphics[scale=0.87]{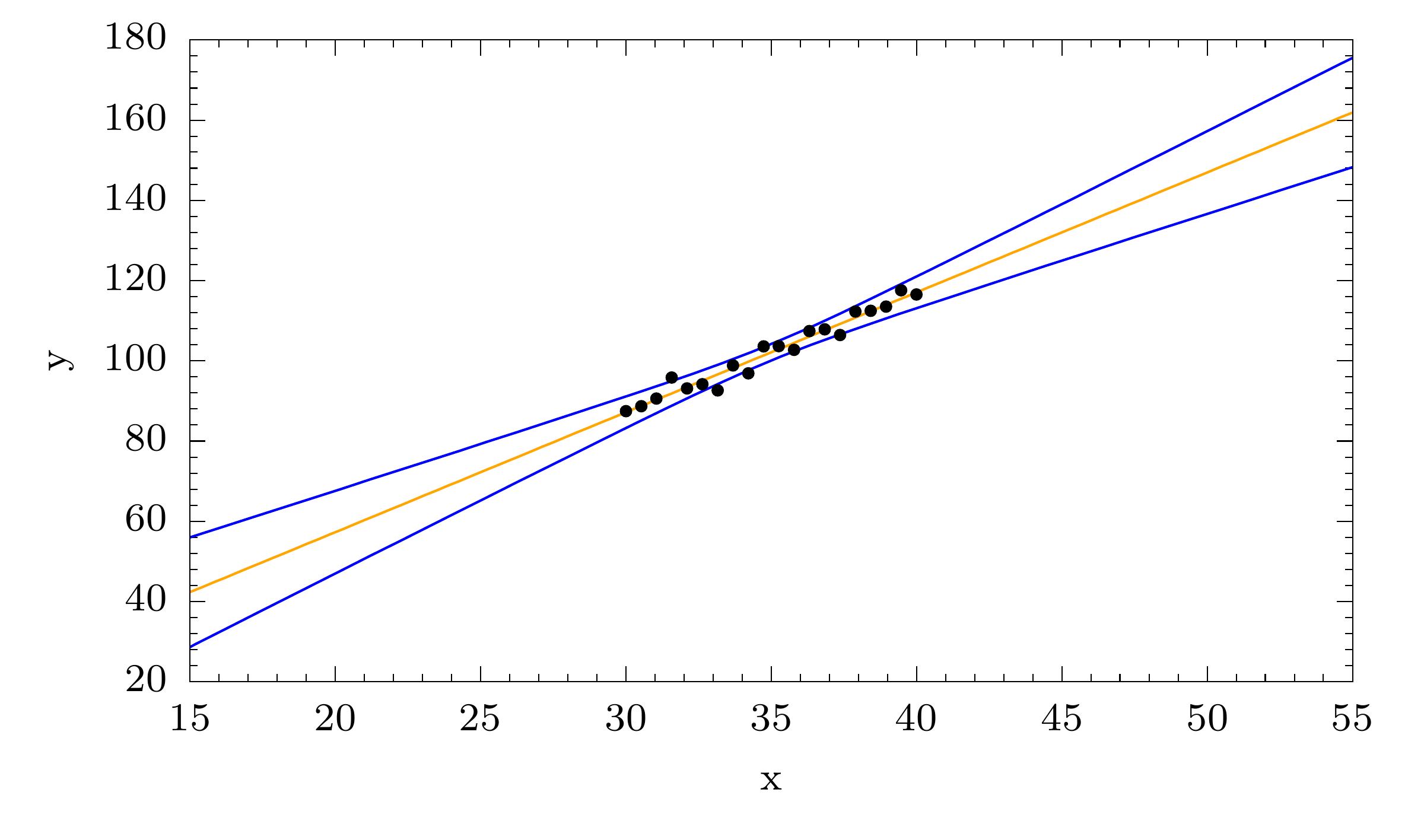}
      \caption{\textbf{Prediction uncertainty for linear models.} Black dots are the arbitrary data fitted with a linear model and generated by adding random Gaussian noise to 20 values of y sampled from the function $y=3x-2$. The best-fit line is reported in orange. The two blue lines corresponds to $\hat{y} \pm 5s$ (see Eq. \ref{error_prediction}).}
      \label{plot_prediction_error}
\end{figure}

Fig. \ref{plot_prediction_error}, based on Eq. \ref{error_prediction}, is very instructive about how the error is estimated\cite{rotondi2011probability}

\begin{itemize}
    \item  the variance associated with a prediction increases as we move away from the centroid of  the distribution of data as compared to the variance of its distribution, due to the presence of $(x-\left\langle x  \right\rangle)^2 / \bar{s}_x^2$;
    \item it is bounded from below from how well the linear regression model fits the preexisting data.
\end{itemize}

In general, the input $\textbf{x}$ has arbitrary dimension $p \times 1$ and Eq. \ref{2d-force_field} has to be written as
\begin{equation}
    Y=\textbf{x}^{T}\textbf{c} + Z \:,
\end{equation}
where \textbf{c} is a vector of coefficients as in Eq. \ref{SNAP_vectorial}. \\

We report here the generalization of Eq. \ref{error_prediction} for the variance of the prediction for a multidimensional input
\begin{equation}
    s^2=s_z^2[1+\textbf{x}^{T}(\lambda \textbf{I}+\textbf{X}^{T}\textbf{X})^{-1}\textbf{x}] \:,
\label{s_error}
\end{equation}
where $\textbf{X}$ is the matrix defined in Eq. \ref{loss_function} and 
\begin{equation}
s_z^2= \frac{(\textbf{X} \textbf{c} -\textbf{Y})^{T}(\textbf{X} \textbf{c} - \textbf{Y}) + \lambda\left\| \textbf{c} \right\|^2}{n-p-1}\:.
\label{s_z_definition}
\end{equation}
The quantity $n$ appearing in Eq. \ref{s_z_definition} is the number of labelled data in the dataset plus the number of equations corresponding to regularization and other constraints, while \textbf{Y} is the same as in Eq. \ref{loss_function}.
The relation in Eq. \ref{confidence_interval} is still valid, but now $t_{1-\alpha}$ is the quantile of the \textit{t-Student} distribution with $n-p-1$ degrees of freedom, where $p$ is the number of parameters to be estimated.\\

If we want to weight differently specific subsets of data, e.g. in case different weights were given to forces and energies when both are used to train the model, we can introduce the transformed variables $\tilde{\textbf{Y}}=\textbf{W}^{\frac{1}{2}}\textbf{Y}$ and  $\tilde{\textbf{x}}=\textbf{W}^{\frac{1}{2}}\textbf{x}$, where $\textbf{W}$ is a $N_{data} \times N_{data} $ diagonal matrix with the square roots of the weights on the diagonal, i.e. $\textbf{W}=diag(1,1,\dots{},w,w,\dots{})$. By definition we fix the weights for the forces equal to 1 and energies are weigthed by the factor $w$. The linear regression then takes the form
\begin{equation}
    \tilde{Y}=\tilde{\textbf{x}}^{T}\textbf{c} + Z \:.
\end{equation}
 Defining $F_{j}$ as the force acting on an atom in the system along a certain Cartesian direction (index $j$ runs both on atoms and Cartesian coordinates, e.g. $F_{1}$ acts on atom $1$ along the $x$-axis, $F_{2}$ acts  on atom $1$ along the $y$-axis and so on) and $E$ as the energy of a given configuration, the loss function takes the following form
\begin{equation}
    \begin{split}
   \mathcal{L}=\frac{\sum_{j}^{N_{data}}[ \sum_{i}^{3N_i} (F_{i}^{DFT}-F_{i}^{ML})_{j}^{2}}{2} +\\
+  \frac{ w(E^{DFT}-E^{ML})^{2}_{j}]+\lambda\textbf{c$^{T}$}\textbf{c}}{2}  \:.  
    \end{split}
    \label{mixed_l}
\end{equation}
Eqs. \ref{s_error} and \ref{s_z_definition} then become
\begin{equation}
    s^2=s_z^2\left[ \frac{1}{w}+\textbf{x}^{T}(\lambda \textbf{I}+\textbf{X}^{T}\textbf{W}\textbf{X})^{-1}\textbf{x} \right] \:,
    \label{s_error_weights}
\end{equation}
and
\begin{equation}
    s_z^2= \frac{(\textbf{X} \textbf{c} -\textbf{Y})^{T}\textbf{W}(\textbf{X} \textbf{c} - \textbf{Y}) + \lambda\left\| \textbf{c} \right\|^2}{n-p-1}\:,
    \label{s_z_error}
\end{equation}
respectively.\\

Now that we have defined a rigorous way to estimate uncertainties on the ML model predictions through Eqs. \ref{s_error} and \ref{s_error_weights}, we are ready to discuss how these quantities inform the AL protocol. \\

A conventional approach to AL would require running a new electronic structure calculation every time the condition 
\begin{equation}
 s > k_{thresh}
 \label{criterion_stop}
\end{equation}
is achieved, where $s^2$ is the predicted variance of residuals and $k_{thresh}$ is a static user-defined threshold corresponding to the desired accuracy. In this work we explore a dynamical definition of $k_{thresh}$, such as
\begin{equation}
    k_{thresh}=\delta \cdot s_z \:, 
    \label{thresh}
\end{equation}
where $s_z$ is the square root of the variance of the residuals for the training set calculated as in Eq. \ref{s_z_definition}, and $\delta$ is set by the user. Setting such a dynamic threshold effectively allows to decouple the definition of stopping criterion for AL from the error on the training set. Indeed, $k_{thresh}$ then becomes identical to the square root of the quantities in square bracket in Eqs. \ref{s_error} and \ref{s_error_weights}, which are independent on $s_z$ and are bounded from below to the value of 1. This approach has the advantage to avoid that AL stops too frequently in case the error on the training set increases as new structures are included in it, and to make the definition of $k_{thresh}$ exportable across different systems.\\

The implementation of Eq. \ref{criterion_stop} is trivial when the uncertainty on energies is the only quantity evaluated. In such case Eq. \ref{s_error} simply output a scalar quantity. However, in the case of training on forces, $\textbf{x}$ in Eq. \ref{s_error} and Eq. \ref{s_error_weights} is a $3N_{at} \times 3N_{at}$ matrix, because we are simultaneously predicting $3N_{at}$ forces components. The output matrix is the covariance matrix for the new prediction, where the diagonal elements represent the variances of the predictions on the single forces. In such case Eq. \ref{criterion_stop} is implemented by taking the largest value of the diagonal elements of the matrix $s^2$.

\subsection*{Connections with Bayesian uncertainty prediction}

Let us now briefly show the similarities between the method just outlined and the Bayesian approach reported in ref. \cite{bishop2006pattern}. \\

In a Bayesian framework, a distribution \textit{a priori} for the parameters must be defined, often taken as an isotropic Gaussian
\begin{equation}
    p(\textbf{c})=\mathcal{N}(\textbf{0},\alpha^{-1}\textbf{I})\:.
    \label{gauss}
\end{equation}
In Eq. \ref{gauss}, $\alpha$ measures the spread of the parameters around the mean and it is assumed to be equal to the identity matrix \textbf{I} for all the parameters.
An \textit{a posteriori} distribution can thus be obtained by combining the \textit{a priori} distribution with the likelihood function. The \textit{a posteriori} distribution is again a Gaussian function with mean \textbf{m$_N$} and covariance \textbf{S$_N$}
\begin{equation}
p(\textbf{c}| \textbf{X},\textbf{Y})=\mathcal{N}(\textbf{m$_{N}$},\textbf{S$_{N}$})\:,
\label{a_posteriori}
\end{equation}
\begin{equation}
\textbf{m$_N$}=\beta\textbf{S$_{N}$}\textbf{X}^{T}\textbf{Y}\:,
\label{map_bayes}
\end{equation}
\begin{equation}
\textbf{S$_{N}^{-1}$}=\alpha\textbf{I}+\beta\textbf{X}^{T}\textbf{X}\:,
\end{equation}
where $\beta$ is the inverse of the variance of $Z$. It can be shown \cite{bishop2006pattern} that the maximization of the logarithm of the \textit{a posteriori} distribution in Eq. \ref{a_posteriori} is equivalent to the problem of minimizing the loss function in \ref{loss_function} with $\lambda=\alpha/\beta$ and that Eq. \ref{map_bayes} is equivalent to Eq. \ref{classical_result}.
Given the \textit{a posteriori} distribution, we can finally obtain the predictive distribution to make predictions $y_*$ on unlabeled input \textbf{x}$_*$
\begin{equation}
    p(y_*| \textbf{x}_*,\textbf{Y},\alpha,\beta)=\mathcal{N}(\textbf{m$_{N}$}^{T}\textbf{x}_*,\sigma^2_N(\textbf{x}_*))\:,
\end{equation}
where $\sigma^2_N(\textbf{x}_*)$ is given by
\begin{equation}
    \sigma^2_N(\textbf{x}_*)=\frac{1}{\beta}+\textbf{x}_*^{T}\textbf{S}_N\textbf{x}_*\:.
    \label{bayesian_variance}
\end{equation}

By setting $\lambda=\alpha/\beta$, the expression of the variance in Eq. \ref{s_error} becomes equivalent to the variance of the prediction in the Bayesian framework in Eq. \ref{bayesian_variance}. In the latter approach the values of $\alpha$ and $\beta$ are obtained by maximizing the evidence function \cite{bishop2006pattern}. \\

An AL method for linear MLFFs exploiting Eq. \ref{bayesian_variance} has recently be reported by Oord et al. \cite{Csany_HAL}. The key difference with our implementation lies in the fact that SNAP makes it feasable to use Eq. \ref{s_error} as is, while in the work of Oord et al. \cite{Csany_HAL} Eq. \ref{bayesian_variance} had to be approximatively estimated due to the large number of unknown parameters in their model. 

\subsection*{Connections with D-optimality design}

As done in the last section, we here want to briefly unravel the similarities between the approach presented in this work and the one proposed by Podryabinkin et al. \cite{PODRYABINKIN2017171} and based on the concept of D-optimality.\\

Given a pool of unlabeled data, the D-optimality criterion states that the optimal selection of points to label is the one that maximizes the determinant of \textbf{X}$^{T}$\textbf{X}\cite{D-optimal_design}.
To make this principle appealing for an on-the-fly AL procedure we have to quantify how much the determinant of  \textbf{X}$^{T}$\textbf{X} changes when a new unlabeled configuration is added to the training set.
If we indicate with \textbf{X}$^{'}$ the matrix including the preexisting training set with the addition of the new point \textbf{x}$^{*}$, then
\begin{equation}
\text{det}(\textbf{X}^{'T}\textbf{X}^{'})= \text{det}(\textbf{X}^T\textbf{X})\cdot \left [ 1+d(\textbf{x}^*) \right]\:,
    \label{D_opt_change_determinant}
\end{equation}
where
\begin{equation}
    d(\textbf{x}^*)=\textbf{x}^{*T}(\textbf{X}^{T}\textbf{X})^{-1}\textbf{x}^{*} \:.
    \label{D_opt_d_term}
\end{equation}
The term in Eq. \ref{D_opt_d_term} can be suitably rewritten in case of regularization or in presence of a weight matrix and shown to be equivalent to the second term in Eqs. \ref{s_error} and \ref{s_error_weights}.
If we set a dynamic threshold
\begin{equation}
    k_{thresh}=\delta\cdot\text{det}(\textbf{X}^T\textbf{X}) \:,
\end{equation}
an \textit{ab initio} calculation is triggered when 
\begin{equation}
    \left [ 1+d(\textbf{x}^*) \right] > \delta \:.
    \label{criterium_D_optimum}
\end{equation}
It can be easily shown that the criterion to stop AL in Eq. \ref{criterium_D_optimum} is equivalent to the one in Eq. \ref{criterion_stop}. At the best of our knowledge, this connection between linear regression uncertainty prediction and D-optimality has never been established before in the context of AL for linear MLFFs.

\section*{Results}

\subsection*{Learning the set rMD17 of ab initio molecular dynamics trajectories}

\begin{table*}[t]
\caption{\textbf{Root mean square error on training and test sets for trajectories in the rMD17 dataset.} Results related to TE (TF) are reported out of (in) parentheses. The RMSE of energies and forces are reported respectively in kcal/mol and kcal/mol/\AA$ $.Cutoff radius values are reported in \AA$ $ and have been chosen to minimize the error on the test error of the energies (TE) or forces (TF). Results are obtained by setting the value of the threshold parameter $\delta$ to 1.5. For every compound, we report three rows corresponding in order to the results obtained with \textit{training AL}, \textit{1000-random} and \textit{N-random}. }
\centering
\begin{tabular}{c c| c c c c}
\toprule
\textbf{Compound}  \hspace{0.1cm} &  \hspace{0.1cm}  \textbf{cutoff}  \hspace{0.1cm} & \hspace{0.1cm} \textbf{TSS} \hspace{0.1cm} & \hspace{0.1cm} \textbf{RMSE Tr} \hspace{0.1cm} & \hspace{0.1cm} \textbf{RMSE Te E} \hspace{0.1cm} & \hspace{0.1cm} \textbf{RMSE Te F}\hspace{0.1cm}\\ \hline

\multirow{3}*{Benzene} &  & 216 (30) & 0.12 (0.86) & 0.12 (0.10) & 1.17 (0.75)  \\
& 4.0 (4.0)& 1000  & 0.08 (0.61) & 0.09 (0.09)  & 0.90 (0.61)  \\
& & 216 (30)  & 0.08 (0.63) & 0.15 (0.10)  & 1.54 (0.69) \\ \hline

\multirow{3}*{Aspirin} &  & 464 (46) & 1.83 (7.53)  & 2.58 (3.06) & 11.27 (7.64) \\
& 3.0 (3.0)& 1000  & 1.95 (7.12) & 2.31 (2.80) & 9.94 (7.04) \\
&  & 464 (46) & 1.71 (6.93)  & 2.63 (2.93)  & 10.89 (7.48) \\\hline
\multirow{3}*{Uracile} &  & 516 (54) & 0.88 (4.25)  & 1.09 (1.06) & 6.61 (4.63) \\
& 3.5 (3.75) & 1000 & 0.74 (3.99) & 0.96 (1.06) & 5.74 (3.96)  \\
& & 516(54) & 0.69 (3.70) & 1.07 (1.09) & 6.40 (4.33)  \\\hline

\multirow{3}*{Naphthalene} &  & 310 (24) & 0.66 (3.18) & 0.65 (0.85) & 3.80 (3.13) \\
& 3.5 (3.5)& 1000 & 0.59 (2.69) & 0.65 (0.73) & 3.29 (2.70) \\
& & 310 (24)& 0.48 (2.41)& 0.76 (0.86) & 3.68 (2.96) \\\hline

\multirow{3}*{Salycilic acid} &  & 473 (48) & 1.34 (6.16)& 1.54 (1.80) & 8.37 (6.23) \\
& 3.0 (3.0)& 1000 & 1.17 (5.11)& 1.38 (1.62) & 7.82 (5.21) \\
& & 473 (48) & 1.03 (4.51) &  1.6 (1.73) & 8.57 (5.99)  \\\hline

\multirow{3}*{Malonaldehyde} &  & 448 (67) & 1.35 (6.10)& 1.53 (1.94) & 8.09 (5.95)\\
& 3.0 (3.0)& 1000 & 1.08 (5.01) & 1.26 (1.81) & 6.84 (5.16) \\
& & 448 (67) & 0.95 (4.55)& 1.53 (2.03) & 8.11 (5.74)  \\\hline

\multirow{3}*{Ethanol} & & 492 (67) & 0.99 (5.02) & 0.99 (1.48) & 5.98 (4.81)\\
& 3.0 (3.0) & 1000 & 0.71 (5.02)& 0.87 (1.01) & 4.97 (4.82)\\
& & 492 (67) & 0.54 (3.45) & 0.99 (1.18)& 5.78 (4.55)\\\hline

\multirow{3}*{Toluene} & & 339 (33) & 1.13 (4.42) & 1.22 (1.55)& 5.63 (4.26)\\
& 3.0 (3.0)& 1000 & 0.92 (3.81)& 1.05 (1.39)& 5.16 (3.84)\\
& & 339 (33) & 0.79 (3.12)& 1.21 (1.50)& 5.73 (4.63)\\\hline

\multirow{3}*{Azobenzene} &  & 365 (38) & 0.91 (3.44) & 1.16 (1.34)& 4.56 (3.37)\\
& 3.5 (3.25) & 1000 & 0.83 (3.11)& 1.01 (1.22)& 3.87 (3.11)\\
& & 365 (38) & 0.71 (2.69)& 1.04 (1.29) & 4.30 (3.35)\\ \hline

\multirow{3}*{Paracetamol} &  & 593 (61)& 1.43 (5.86) & 2.00 (2.11)& 8.21 (5.73)\\
& 3.0 (3.0) & 1000 & 1.36 (5.25)& 1.80 (2.17)& 7.45 (5.24)\\
& & 593 (61) & 1.29 (4.91)& 2.00 (2.07) & 8.15 (5.81)\\ \hline
\\
\label{rMD17_table}
\end{tabular}
\end{table*}

We assess the validity of our method by benchmarking it on the rMD17 dataset. The latter comprises 100k configurations (geometries, energies and forces) sampled from a single trajectory of \textit{ab initio} MD at 500 K for 10 organic molecules of size between 9 and 24 atoms. For all of them, we train one SNAP potential over either energy data (TE) or over forces (TF). For TE, the initial training set, namely the training set before AL starts, includes the first three configurations in the dataset, while for TF only the first structure of the AIMD trajectory is used.\\

We compare results obtained with three different training sets:
\begin{itemize}
    \item \textit{training-AL} built with the AL workflow presented in this work;
    \item \textit{1000-Random} obtained by training the model on 1000 random configurations;
    \item \textit{N-Random} built by training the model on the same number of structures found with \textit{training-AL} but selected randomly.
\end{itemize}
The parameters $\lambda$ and $N_{k}$ (see Eq. \ref{SNAP_equation}) are kept fixed to the values of 0.1 and 56, respectively. We test different values of $R_{cut}$ in the range [3.0, 5.0] \AA$ $ with a step 0.5 \AA$ $ for TE and 0.25 \AA$ $ for TF. In Tab. \ref{rMD17_table} we report only the value of $R_{cut}$ that minimizes the error on training and test sets. The test set is the same for the three MLFFs, and it is made of 1000 configurations randomly selected from each trajectory in the dataset.\\

Results are reported in Tab. \ref{rMD17_table} and show that SNAP achieves a good accuracy on the prediction of energy over both TE and TF, despite having orders of magnitude less degrees of freedom compared to other models\cite{doi:10.1021/acs.jctc.1c00647}.

Interestingly, we obtain comparable results for all the training sets, including \textit{1000-random} and \textit{N-Random}. On the one hand, this demonstrates that a even a few configurations are enough to obtain converged results, proper of large training sets such as \textit{1000-Random}. On the other hand, it is not yet clear how AL would improve over a random selection of the training set. It is important to remark that a comparable level of accuracy does not imply that the FF generated with different datasets lead to MD simulations of comparable quality. The main advantage of using the present method is to guarantee that the uncertainty of energy and/or forces on the configurations sampled by MD is not exceeding a certain value and requiring a minimum number of electronic structure simulations, thus minimizing the computational overheads and the instability of the MD trajectory at the same time.

\begin{figure}
      \centering
      \includegraphics[scale=0.87]{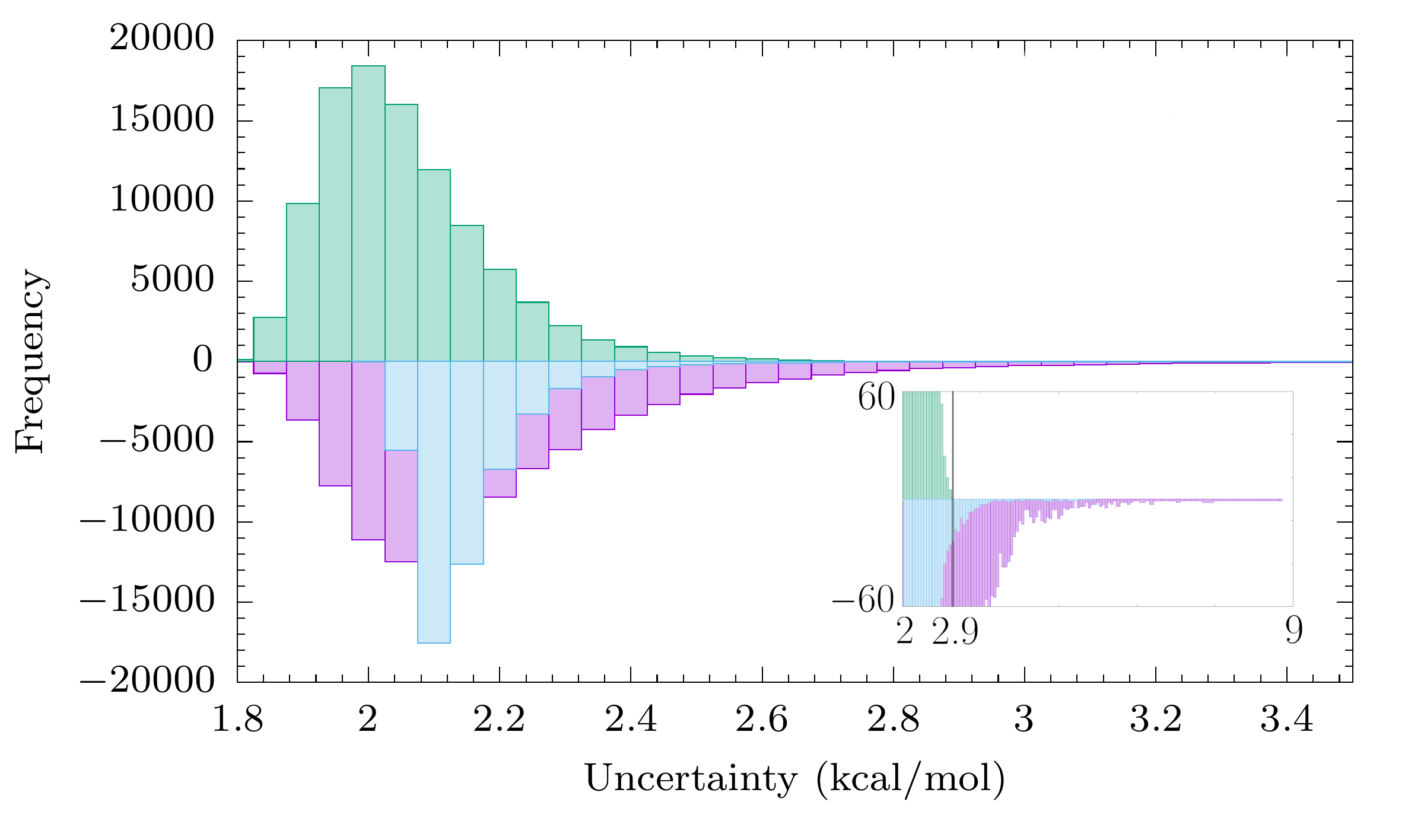}\\
      \includegraphics[scale=0.87]{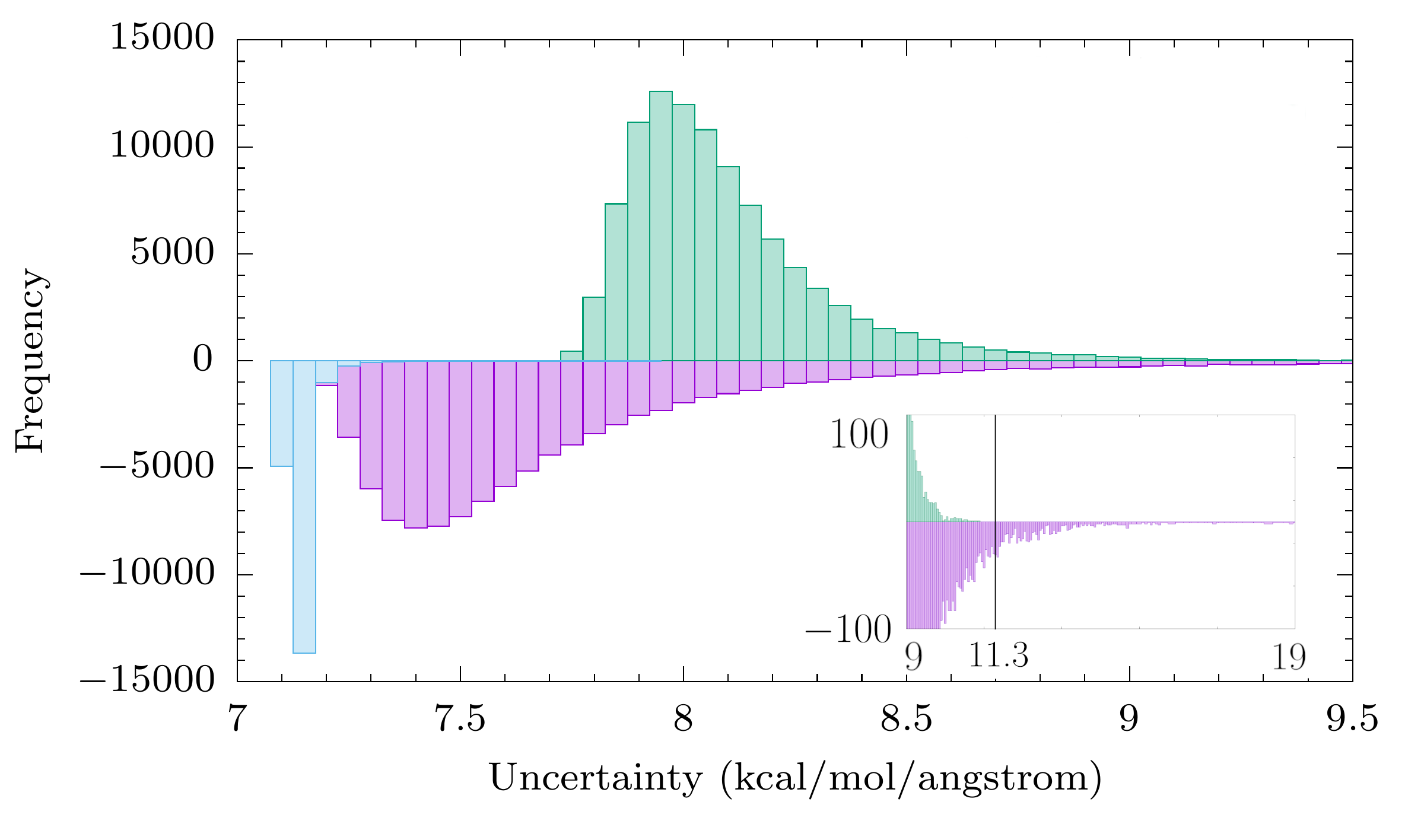}
      \caption{\textbf{Uncertainty distribution of predictions during MD.} The top (bottom) panel reports the distribution of the uncertainty evaluated on the trajectory for aspirin taken from the rMD17 dataset using the TE (TF) method. Results are plotted for the three different training sets, namely \textit{training-AL} in green, \textit{N-Random} in violet and \textit{1000-Random} in cyan. The insets report a zoom over the tail of the distributions for large values of uncertainty. A vertical black line marks the value of $k_{thresh}$ used during AL.}
      \label{distribution_uncertainty_aspirin}
\end{figure}

To prove this point we evaluate the uncertainty during the MD trajectory of aspirin by using the three different training sets and reports results in Fig. \ref{distribution_uncertainty_aspirin}.

Fig. \ref{distribution_uncertainty_aspirin} nicely shows that the tail of the distribution of uncertainty on both energy and forces (top and bottom panel, respectively) has a much longer tail for the \textit{N-Random} with respect to \textit{training-AL}. This fact directly translates into a minimization of the probability that critical configurations are sampled during MD, where the error on the predicted forces is so large to potentially lead the simulation astray. 
As emphasized by the inset of the top panel of Fig. \ref{distribution_uncertainty_aspirin}, training the model over a large values of energies, i.e. for the \textit{1000-Random} set, does not overcome this issue and the tail of the distribution of uncertainty during MD exceeds the one achieved with AL. 

The same qualitative results are obtained for the training over forces (bottom panel of Fig.  \ref{distribution_uncertainty_aspirin}), with the only difference that in this case the uncertainty achieved over the set \textit{1000-Random} is dramatically suppressed. This is in agreement with the fact that such training set contains a large volume of information coming from $3N_{at}$ values of forces for each MD frame, therefore largely exceeding the amount of data available in the other two sets. Interestingly, this result further demonstrates that the error over a training/test set is not an sufficient indicator of the robustness of a force field. Indeed, for the training with forces, all sets achieve similar RMSE values but perform quite differently in terms of uncertainty over predictions. \\

\subsection*{Boot-strapping of machine-learning force fields with active learning}

\begin{figure}[h!]
      \centering
      \includegraphics[scale=0.35]{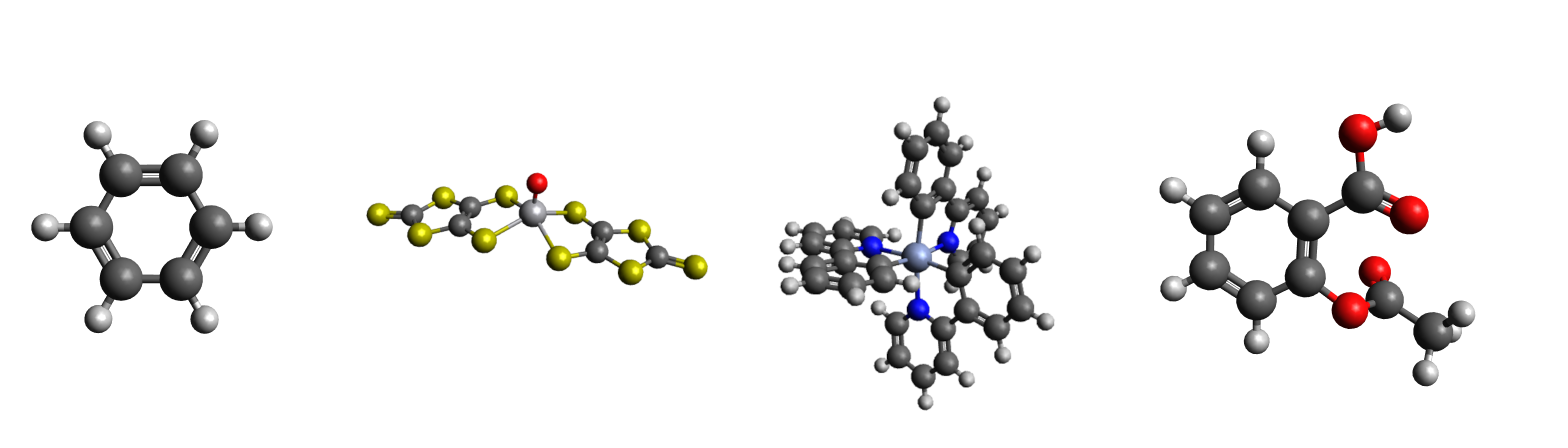}
      \caption{\textbf{Molecular structure of the four benchmark molecules.} From left to right: benzene, VO(dmit)$_2$ , Cr(ppy)$_{3}$ and aspirin. Colour code: oxygen in red, carbon in dark grey, sulphur in yellow, nitrogen in blue, hydrogen in white, vanadium in light grey and chromium in light blue.}
      \label{molecule_to_study}
\end{figure}

The tests over the rMD17 dataset provides important insights on the ability of the proposed AL strategy to achieve a well-balanced training set with a minimal number of configurations. However, this analysis does not address some additional crucial challenges connected with the generation of MLFFs. Firstly, performing the training on pre-compiled datasets, such as rMD17, does not take into account the challenge of selecting realistic configurations to be added to the training set in the first place. Indeed, all the structures contained in the rMD17 set are realistic by construction, having been generated by AIMD. However, this situation does not correspond to common realistic scenarios, where the possibility to run AIMD, even if short, would at least partially defeat the purpose of generating a MLFF. Alternative ways to kick-start the generation of a MLFF with AL have been explored. One simple but often inefficient approach consists in generating random atomic distortions, while displacing molecules along normal mode coordinates has also been tested. Alternatively, if another force field is available, one can use it in place of ab initio methods to generate a first conformational sampling.\cite{Csany_HAL}. Although widely used, pre-sampling methods often lead to a biased training set, thus posing limits to either the accuracy or the stability of the final MLFF.\\

In this section, we show that the proposed scheme is able to achieve the efficient training of a robust MLFF starting from the sole equilibrium configuration of a molecular compound. We perform simulation for four molecules of growing complexity, whose structure is reported in Fig. \ref{molecule_to_study}: benzene, aspirin, VO(dmit)$_2$ (where dmit=1,3-dithiole-2-thione-4,5-dithiolate), and Cr(ppy)$_{3}$ (where Hppy = 2-phenylpyridine). Whilst benzene can be regarded as a toy system, the generation of a MLFF for aspirin already presents some real-scenario challenges connected to its flexibility. On the other hand, VO(dmit)$_2$ and Cr(ppy)$_{3}$ are two coordination complexes with an open shell configuration of interest for the communities of molecular magnetism\cite{VOdmit_reference_paper} and photo-luminescence\cite{Crppy3_reference_article}, therefore rightfully belonging to the class of realistic systems. There is very sparse literature for coordination or magnetic compounds compared to organic molecules and we here provide evidence that the proposed AL scheme is general enough to deal with the inherent complexity of such molecules. \\

The AL protocol is implemented as for the rMD17 set, with the crucial difference that the MLFF is used to propagate the MD from the very beginning, therefore not relying on a pre-compiled trajectory. This is particularly challenging during the early stages of the training, where only very limited information is available to the ML model and the prediction of forces will in general be very poor, potentially leading to catastrophic results. 

We arbitrarily define the AL simulation converged when the algorithm has completed five consecutive MD trajectories of 100 ps without finding new structures. This criterion allows to reinitialize the velocity periodically and enforces an ergodic exploration of the configuration space. As for the rMD17 test, the initial training set is constructed with just three configurations for the training over sole energy values. The latter three configurations correspond to the equilibrium structure and two randomly displaced structures by 0.05 \AA$ $ max. For the training with forces, we instead trained the first MLFF using only information coming from the equilibrium structure. Finally, the test set is constructed by taking 100 configurations sampled every 1 ps from the last MD trajectory explored during AL. MD is performed at 300 K using the thermostat by Bussi et al. \cite{NVT-rescaling}.

The cutoff radius of the $N_k=56$ bispectrum components is set to 4 \AA$ $ for all chemical species and the regularization value $\lambda$ is set to 0.1. All \textit{ab initio} calculations are performed with the software ORCA\cite{doi:10.1063/5.0004608}. For all four systems we employ the PBE functional\cite{PBE_functional}, with the basis set def2-TZVPP and def2/J auxiliary basis for the RI approximation. In the case of benzene, VO(dmit)$_2$ and aspirin D3 vdW corrections are employed.\cite{D3BJ_grimme,Revised_DFT-D3}\\

\begin{figure}
      \centering
      \includegraphics[scale=0.87]{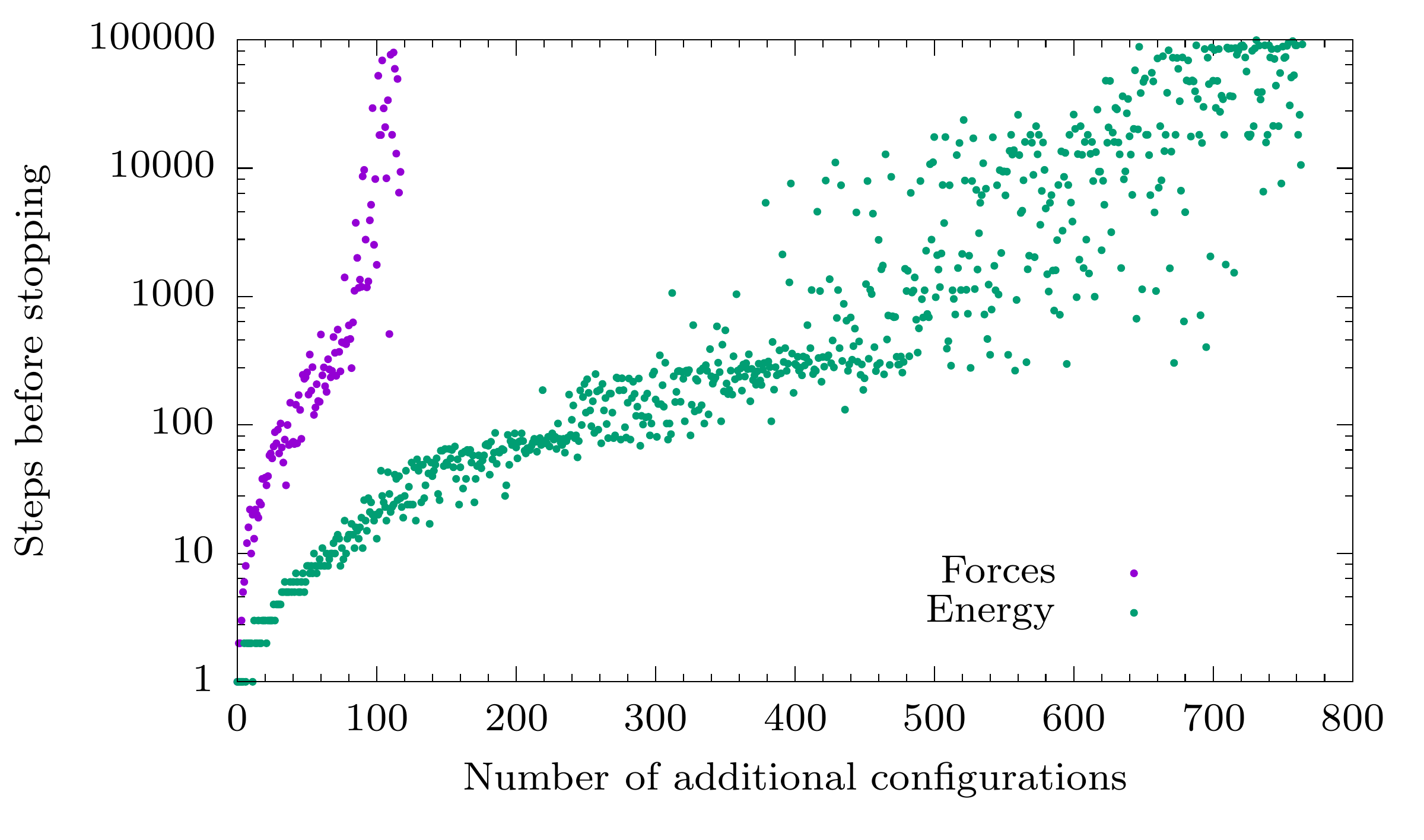}
      \caption{\textbf{Acquisition curve for VO(dmit)$_2$.} The plot shows the number of steps performed during MD in an active learning cycle before finding a configuration to include in the training set for $\delta =1.5$.  Results fro the training on only energy values are reported in green and the ones for a training on forces are reported in violet.}
      \label{learning_rate}
\end{figure}

The results on the root mean square errors (RMSE) and final training set size are shown in Tab. \ref{results_AL_our_compounds}. For benzene and VO(dmit)$_2$, the model achieves chemical accuracy on both training and test set (RMSE $<$ 1 kcal/mol) for every value of $\delta$. Notably, the number of structures required to achieve the generation of a stable and accurate force field is dramatically reduced by training on forces instead of the sole energy values. Fig. \ref{learning_rate} further emphasize this result by reporting the number of MD steps performed before a new DFT calculation is requested by the AL algorithm.
We further test our model on the more challenging Cr(ppy)$_3$ and aspirin. The training with energy once again achieves very good results close to chemical accuracy. Given the higher structural complexity of these two compounds, more structures are needed to converge the AL simulations. However, differently from the previous two compounds, the training on sole forces this time leads to the generation of force fields that reach unphysical configurations during MD. Even using small values of $\delta$, very close to the lowest limit of 1, does not fix the problem. We overcome this apparent limit of the AL algorithm by combining the benefits of training on energies (stability and accuracy) with the benefits of training on forces (convergence rate) by training the model on energies and forces at the same time. We have set $w=9$ and $w=81$ for aspirin and Cr(ppy)$_3$, respectively, in Eq. \ref{mixed_l} and  $w=1$ in Eq. \ref{s_error_weights}. The results reported in Tab. \ref{results_AL_asp_cromo} demonstrate the viability of this approach and show that it is possible to obtain comparable performances over the test set's RMSE by training with either sole energies or energies and forces for complex compounds. Crucially, in the latter case, only a small fraction of ab initio calculations are required.

\begin{table*}[t]
\caption{\textbf{RMSE on training and test sets for four selected compounds.} The RMSE of energies and forces are reported in kcal/mol and kcal/mol/\AA$ $, respectively, for the different sets, i.e. training (Tr) and test (Te), and for training performed on either energy (out of parentheses) or force values (in parentheses).
The training set size (TSS) selected by the active learning algorithm is also reported for different values of the threshold parameter $\delta$.}
\centering
\begin{tabular}{c c | c c c c}
\toprule
\textbf{Compound}  \hspace{0.1cm}  &  \hspace{0.1cm}  \textbf{$\delta$}  \hspace{0.1cm}  & \hspace{0.1cm} \textbf{TSS} \hspace{0.1cm} & \hspace{0.1cm} \textbf{RMSE Tr } \hspace{0.1cm} & \hspace{0.1cm} \textbf{RMSE Te E} \hspace{0.1cm} & \hspace{0.1cm} \textbf{RMSE Te F}\hspace{0.1cm}\\\hline
%\textbf{RMSE Tr F}\hspace{0.1cm} & \hspace{0.1cm}\textbf{RMSE Te F}\hspace{0.1cm} & \hspace{0.1cm}\textbf{RMSE Te E} \hspace{0.1cm}\\ 
 %&  & \multicolumn{3}{c}{\textbf{Training on Energies}} & \multicolumn{4}{c}{\textbf{Training on Forces}} \\ \hline

\multirow{4}*{Benzene} & 1.5 & 387 (57) & 0.09 (0.68) & 0.09 (0.1) & - (0.66) \\
&1.75 & 260 (40) & 0.11 (0.6)& 0.08 (0.3)& - (0.52)\\
&2.0 & 187 (31) & 0.08 (0.58)  &  0.09 (0.07) & - (0.68)\\
&2.25 & 157 (26) & 0.09 (0.56) & 0.11 (0.09) & - (0.73)\\ \hline
%\midrule
%\multirow{4}*{Butane} & 1.5 & 947 & 0.55 & 0.76 \\
%& 1.75 & 651 & 0.49 & 1.00 & 38 & 2.92 & 4.95 & 1.68\\
%& 2.0 & 513 & 0.40 & 0.95\\
%& 2.25 & 429 & 0.40 & 0.77\\ \hline

\multirow{4}*{VO(dmit)$_2$} & 1.5 & 768 (121)  & 0.42 (1.11) & 0.53 (0.99) & - (1.17)\\
 &1.75 & 489 (82)  & 0.38 (1.04)& 0.50 (0.70) & - (1.38)\\
 & 2.0 & 381 (73) & 0.41 (1.16) & 0.62 (1.23)&  - (1.24)\\
 & 2.25 & 317 (62) & 0.36 (1.09) & 0.81 (0.84) & - (1.30)\\ \hline
%\midrule

\multirow{4}*{Aspirin} & 1.5 & 1319 (-)  & 1.08 (-) & 1.76 (-) & - (-)\\
& 1.75  & 955(-)  & 1.25 (-) & 4.20(-) & - (-)\\
 & 2.0 & 735(-) & 0.89 (-) & 2.69(-) & - (-)\\
 & 2.25  & 617 (-) & 0.89 (-) &  2.77(-) &  - (-)\\ \hline
\multirow{4}*{Cr(ppy)$_3$} & 1.5 & 1656 (-) & 0.79 (-) & 1.9 (-) & - (-)\\
&1.75 & 1172 (-) & 0.90 (-) & 2.18 (-) & - (-)\\
&2.0  & 896 (-) & 0.73 (-)& 1.62 (-)& - (-) \\
&2.25  & 779 (-) & 0.84 (-) & 2.79 (-) & - (-) \\ \hline
\\
\label{results_AL_our_compounds}
\end{tabular}
\end{table*}

\begin{table*}[t]
\caption{\textbf{RMSE on training and test sets for aspirin and Cr(ppy)$_3$.} The RMSE of energies and forces are reported in kcal/mol and kcal/mol/\AA$ $, respectively, for the different sets, i.e. training (Tr) and test (Te), and for training performed on energy (out of parentheses) and force values (in parentheses).
Training in this case has been done both on energies and forces.The relative weight (energies:forces) when solving the regression problem is 3:1 for aspirin and 9:1 for Cr(ppy)$_3$. The training set size (TSS) selected by the active learning algorithm is also reported for different values of the threshold parameter $\delta$.}
\centering
\begin{tabular}{c c | c c c c}
\toprule
\textbf{Compound}  \hspace{0.1cm}  &  \hspace{0.1cm}  \textbf{$\delta$}  \hspace{0.1cm}  & \hspace{0.1cm} \textbf{TSS} \hspace{0.1cm} & \hspace{0.1cm} \textbf{RMSE Tr } \hspace{0.1cm} & \hspace{0.1cm} \textbf{RMSE Te E} \hspace{0.1cm} & \hspace{0.1cm} \textbf{RMSE Te F}\hspace{0.1cm}\\\hline

\multirow{4}*{Aspirin}
& 1.3  &  92 & 4.20 (6.64) & 2.60  & 8.52\\ 
& 1.4  &  76 & 4.83 (7.90) & 3.02  & 9.80\\
& 1.5  &  60 & 3.78 (6.50) & 3.57  & 9.15\\
& 1.75 & 53 & 4.92 (8.93) & 3.31  & 10.57\\
\hline
\multirow{4}*{Cr(ppy)$_3$} &1.2 & 148 & 1.57(3.22) & 2.52 & 4.53  \\ 
& 1.3 & 105 & 1.25(3.54) & 2.14 & 4.46\\
& 1.5 & 74 & 1.77(4.64) & 6.90 & 5.86\\
&1.75 & 51 & 1.73(4.03) & 3.69 & 5.10 \\
\hline
\\
\label{results_AL_asp_cromo}
\end{tabular}
\end{table*}
%commentiamo i risultati ottenuti

\section*{Discussion and Conclusions}

The use of machine learning to map the PES of chemical compounds has revolutionized the field of materials modelling, opening up the possibility to simulate nm-sized systems over extended time scales and to sample extremely large portions of the chemical space \cite{Review_Unke}. Since the inception of the field, several different approaches have guided the development of new MLFF frameworks. Important achievements have been reached in the development of elaborated ML models able to fit the PES of chemical compounds with extraordinary accuracy, including for instance long range and non-local interactions\cite{Physnet,Schnet,unke2021spookynet}. Moreover, certain MLFF frameworks have been shown to be able to learn the PES of entire classes of compounds and to generalize to molecules not included in the training set \cite{ANI_Isayev,ANI-2x,chen2022universal}. 

In this contribution, instead, we focused on a different approach, where training robustness and efficiency are valued at the same level of accuracy, at the expense of transferability. We believe that this approach is also required to fulfil all the needs of the MLFFs community. Indeed, given the complexity of the chemical space, we are still far away from having a universally accurate and robust MLFF able to predict the PES of any molecular system, and the application of MLFFs to new chemical systems often requires the generation of \textit{de-novo} dedicated training sets. Such a computationally-demanding task must be dealt with as efficiently as possible in order for MLFFs to become a standard computational tool.

Whilst advanced MLFF frameworks are able to accurately map the PES of relatively simple organic molecules, their training is quite nuanced and often computationally expensive. Moreover, no evidence is yet available on their application to complex compounds with many chemical species. On the other hand, transferable force fields able to predict the PES of general organic compounds are now available, but only for a small number of ethero-atoms\cite{ANI-2x}, and with the important exclusion of coordination compounds of transition metals and rare earths. The latter are key for the simulation of bio-inorganic system, luminescent sensors, catalysts, etc.

Here we have shown that linear models, once combined with an uncertainty-aware active learning strategy, are able to accurately approximate the PES of complex chemical systems with only a handful of electronic structure calculations and without requiring a, often biased, pre-sampling of the conformational space. These key features make it possible to readily train a MLFF for a new compound in a very short amount of human and computational time. Importantly, we have demonstrated that the resulting MLFFs are able to withstand MD at room temperature, which we advocate it should be introduced as a key metric to assess the quality of a MLFF. 

It is also important to remark that the method outlined in this work employs only three hyper-parameters, namely the cutoff radius of the bispectrum components, $R_{cut}$, the relative weight of energies and forces, $w$, and the active learning threshold, $\delta$. Chemical intuition naturally guides the choice of an optimal $R_{cut}$, while tests suggest that excessive fine-tuning of the other two hyper-parameters is not required. Having just a few, not too sensitive hyper-parameters is a key aspect of an efficient and robust MLFF framework, as it makes the model more user friendly and potentially compatible with high-throughput and automated workflows. \\

Several avenues of future development can be envisioned. First and foremost, an in-depth study on the dependency of the MLFFs' accuracy and stability on the choice of the atomic environments' descriptors is required. Here, we implemented our linear ML model with bispectrum components as atomic environments' fingerprints, which we believe offer some advantages. For instance, bispectrum components provide quite a compact description of atomic environments and their number scales linearly with the number of atomic species. Throughout this work, we have used 55 bispectrum components per chemical element, thus never exceeding a total number of adjustable parameters of 224. On the one hand, this small number of descriptors allowed us to generate accurate MLFFs with only a small number of reference ab initio data. On the other hand, a descriptor with only a few degrees of freedom poses a limitation on the accuracy that can be reached by increasing the training set size. Other descriptors such as ACE, the ones used in Moment Tensor and Jacobi-Legendre potentials\cite{Shapeev_MTP,domina2022jacobilegendre}, have been used as building block for linear MLFFs, and might offer a better trade off between accuracy and robustness and merit further investigation.

Another aspect that will require further work concerns the exploration of the limits of linear ML models and AL to deal with a varied number of chemical systems. Although in this work we focused on gas-phase molecular systems, the method should readily apply to condensed matter systems just as well, provided long-range interactions are included in the model. The inclusion of electrostatic and dispersion interactions into MLFFs frameworks has recently received large interest and several promising schemes are now available \cite{c60_vdW,zhang2022deep,behler2021four}. 

The method explored in this work is also promising in terms of the types of chemical properties that can be predicted. Indeed, an equivariant version of linear MLFFs based on SNAP have been recently proposed for the mapping of tensorial properties \cite{annie_tensors}, and the method discussed here readily applies to that scenario, thus further extending the scope of the present work.\\

In conclusion, we have here presented an AL protocol for linear machine learning models able to produce accurate results for complex molecular systems with a minimal number of \textit{ab initio} data and requiring minimal human intervention. We applied this strategy together with the model SNAP and tested its performances over the rMD17 dataset and on the generation of FFs from scratch with no preexisting dataset.
This method successfully leads to force fields able to withstand accurate MD at room temperature with only tens of training configurations, thus paving the way to the automatic and efficient generation of MLFFs for challenging chemical systems.  

\vspace{0.2cm}
\noindent
\textbf{Acknowledgements and Funding}\\
This project has received funding from the European Research Council (ERC) under the European Union’s Horizon 2020 research and innovation programme (grant agreement No. [948493]). Computational resources were provided by the Trinity College Research IT and the Irish Centre for High-End Computing (ICHEC).

\vspace{0.2cm}
\noindent
\textbf{Conflict of interests}\\
The authors declare that they have no competing interests.

%\bibliographystyle{naturemag}
%\bibliography{biblio}

\end{document}